\documentstyle[prl,aps,epsf,floats]{revtex}
\input epsf
\epsfverbosetrue
\begin{document}
\draft
\twocolumn[\hsize\textwidth\columnwidth\hsize\csname  
@twocolumnfalse\endcsname
\title{Pseudogap effects induced by resonant pair scattering}
\author{Boldizs\'ar Jank\'o, Jiri Maly and K. Levin}
\address{The James Franck Institute, The University of Chicago, 5640
S. Ellis Avenue, Chicago IL 60637}
\date{\today}
\maketitle
\begin{abstract}
We demonstrate how resonant pair scattering of correlated electrons 
above $T_c$ can give rise to pseudogap behavior. This resonance in 
the scattering T-matrix appears for superconducting interactions 
of intermediate strength, within the framework of a simple fermionic model.  
It is associated with  a splitting of the single peak in 
the spectral function into a pair of peaks separated by 
an energy gap. Our physical picture is contrasted 
with that derived from  other T-matrix schemes,
with superconducting fluctuation effects, and with preformed pair 
(boson-fermion) models. Implications for photoemission and 
tunneling experiments in the cuprates are discussed. 
\end{abstract}
\pacs{\rm PACS numbers:  74.20.Mn, 74.25.-q, 74.25.Fy, 74.25.Nf,
74.72.-h
\hfill {\tt cond-mat/9705144}}
]
\makeatletter
\global\@specialpagefalse
\def\@oddhead{REV\TeX{} 3.0\hfill Levin Group Preprint, 1997}
\let\@evenhead\@oddhead
\makeatother

It has become clear in recent years that the presence 
of a pseudogap above the superconducting transition temperature,
$T_c$, is a robust feature of the underdoped cuprates. 
This phenomenon is manifested  in thermodynamic \cite{thermo}, 
magnetic \cite{magn}, and
angle-resolved photoemission spectroscopy (ARPES) data 
\cite{arpes}. 
These ARPES experiments, which have established the presence of a
Luttinger volume Fermi surface, 
place important constraints on any pseudogap scenario:
they indicate that the pseudogap
appears directly in the spectral
function and its magnitude and symmetry\cite{arpes} seem to
evolve smoothly into that of the superconducting state. 
Furthermore, the minimum gap points in  the
pseudogap regime retrace the normal state Fermi surface\cite{ding}.

A variety of theoretical scenarios have been proposed,
for the origin of the pseudogap. Quantum 
Monte Carlo simulation studies have been carried out 
on both positive and negative U Hubbard models \cite{QMC}.  
Alternative models relate the pseudogap to either magnetic pairing of 
spins \cite{magnetic}, RVB-like pairing of chargeless 
spinons \cite{RVB}, or precursor superconductivity effects 
\cite{Kivelson}.
The present paper addresses this last 
scenario, in part because of constraints from ARPES data and 
in part, because the cuprates are short coherence length, 
quasi-two dimensional superconductors, with anomalously low 
plasma frequencies\cite{Kivelson,Maly}. They are, therefore, expected to 
exhibit important deviations from an abrupt, BCS-like transition.

In our physical picture, we associate an important component of the 
cuprate pseudogap with {\it resonant} 
scattering between electrons of opposite spin and small total 
momentum. This resonance arises in the presence of intermediate 
coupling and a sizeable Fermi surface. A depression in the 
density of states occurs because states near this Fermi surface
are unavailable for electrons in the Fermi sea to scatter into;
such states are otherwise occupied  by relatively long lived
(metastable) electron pairs. The related  suppression in the 
spectral weight differs from that derived from conventional low 
frequency and long wavelength fluctuation effects \cite{fluct}. 
In the present  case it is the strength of the attractive interaction,
rather than the critical slowing down (in proximity to  $T_c$), 
which leads to the long-lived pair states.  It should be noted 
that our resonant scattering  approach is to  be distinguished 
from previous precursor superconductivity models associated with 
either preformed pairs\cite{boson-fermion} or dynamic 
phase fluctuations\cite{Kivelson}. 

Our starting point is a scheme which connects the strong coupling, 
short coherence length description  of superconductivity formulated 
by Leggett and Nozi\`eres and Schmitt-Rink \cite{NSR} 
with a  well established T-matrix formalism designed to treat normal 
state fluctuation effects in conventional superconductors
\cite{fluct}.  We consider a generic model Hamiltonian
\begin{eqnarray}
{\cal H} & = & \sum_{{\bf k}\sigma} \epsilon_{\bf k} 
c^{\dag}_{{\bf k}\sigma} c^{\ }_{{\bf k}\sigma}
\nonumber \\
& & + \sum_{\bf k k' q} V_{\bf k, k'} 
c^{\dag}_{{\bf k}+{\bf q}/2\uparrow} 
c^{\dag}_{-{\bf k}+{\bf q}/2\downarrow} 
c^{\ }_{-{\bf k'}+{\bf q}/2\downarrow} 
c^{\ }_{{\bf k'}+{\bf q}/2\uparrow},
\label{Hamiltonian}
\end{eqnarray}
where $c^\dagger_{{\bf k}\sigma}$ creates a particle in the 
momentum state ${\bf k}$ with spin $\sigma $, and 
$\epsilon _k =  k^2/2m - \mu$ (we take $\hbar=k_B=1$).
Here  $V_{\bf k,k'} = 
g \varphi_{\bf k}\varphi_{\bf k'}$, where\cite{k0}
$\varphi_{\bf k} = (1+k^2/k^2_0)^{-1/2}$, and $g<0$ is the coupling 
strength. While we consider the $s$--wave symmetry case, $d$--wave 
symmetry can be readily introduced via $\varphi_{\bf k} \rightarrow
(\cos k_x - \cos k_y )$. 
In the scheme of Nozi\`{e}res and Schmitt-Rink  the transition
temperature, $T_c$,
must be obtained in combination with the chemical potential, $\mu$, 
by use of  the Thouless criterion,
$T^{-1}_{\bf q=0}(\Omega=0) = 0$ (see below), and the 
usual equation for particle number. 
These authors have demonstrated that, when the parameter $g$
is varied, the appropriate coupled equations lead 
to an interpolation scheme which contains the BCS 
limit for small $g/g_c$ , where $\mu \approx E_{\rm F}$,
and that of Bose-Einstein condensation in strong coupling,
where $\mu$ becomes negative. Here $g_c = -4\pi/mk_0$ is the
critical value of the coupling at which 
a bound state of the isolated pair first appears. 

\begin{figure}
\narrowtext
\epsfxsize=3in 
\epsfbox{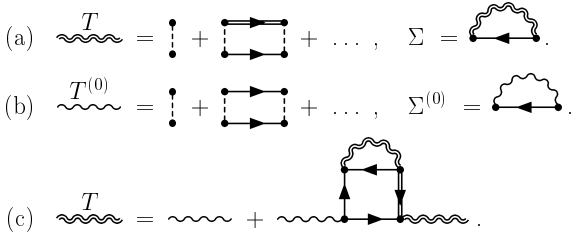}\vspace{-0.1in}
\caption{Diagrams for coupled $\Sigma$, $T$, in full scheme (a) and 
lowest order conserving scheme (b) used here. Fig. 1(c) represents a 
rewriting of $T$ in Fig.1(a).}
\label{fig:self_e}
\end{figure}

In this paper, we extend this formulation so as to  provide a basis for 
computing the spectral function and density of states and to 
simultaneously
incorporate  appropriate   conservation laws\cite{serene}. 
To this end, we calculate the single particle self-energy, 
$\Sigma_{\bf k}(\omega)$, and the T-matrix
as shown diagrammatically in Fig.~\ref{fig:self_e}\cite{Patton}.
Analytically, the self-energy corresponds to 
\begin{equation}
\Sigma_{\bf k}(i\zeta_l)  = \frac{1}{\beta}\sum_{{\bf q},\Omega_n}
\varphi^2_{{\bf k-q}/2}
T_{\bf q}(i\Omega_n) G^{(0)}_{\bf q-k}(i\Omega_n-i\zeta_l), 
\label{self-energy}
\end{equation}
and the T-matrix, which may be written in  the
Dyson form shown in Fig.~\ref{fig:self_e}(c), is given by
\begin{equation}
T_{\bf q} (i\Omega_n) = \left[\frac{1}{g} + \frac{1}{\beta} 
\sum _{{\bf p},\zeta _l}
\varphi^2_{\bf p} G_{{\bf p+q}/2}(i\zeta_l) G^{(0)}_{{\bf
p-q}/2}(i\Omega_n - i\zeta _l)\right]^{-1}.
\label{T-matrix}
\end{equation}
Here $G^{-1}_{{\bf k} }(i\zeta _l) = G^{(0) -1}_{\bf k} (i\zeta _l) - 
\Sigma _{\bf k} (i\zeta _l)$ and $G^{(0)}_{{\bf k}} (i\zeta _l) 
= (i \zeta_l  - \epsilon_k)^{-1}$.
Finally, the spectral function is defined as
$A_{\bf k} (\omega ) = -\pi^{-1}{\rm Im}\,[G_{\bf k}
(i\zeta _l \rightarrow \omega +i0)]$;  this leads
to the density of states,
$N(\omega ) = \sum _{\bf k} A_{\bf k}(\omega)$. 

The choice of diagrams to include  in a T-matrix scheme 
has been extensively discussed in the
literature \cite{Patton,Kadanoff,micnas}.  
Our asymmetric choice [Fig.~\ref{fig:self_e}(a)]
-- in which the T-matrix contains one self-energy renormalized 
and one ``bare'' propagator --  builds on the early work of  Kadanoff 
and Martin\cite{Kadanoff}.
When this diagrammatic scheme was applied to conventional 
superconducting fluctuation effects it was shown\cite{Patton} that
the fluctuation gap above $T_c$ smoothly evolved into the  
superconducting gap below $T_c$. Moreover, this approach is 
known\cite{Patton,Kadanoff} to reproduce the 
conventional BCS theory in the appropriate weak coupling limit.
Direct connection can be made to the related theories of Marcelja
\cite{marcelja} and Yamada and
collaborators \cite{yamada} if the full line in the ``box'' (the pair
fluctuation self-energy) of  Fig.~\ref{fig:self_e}(c) is replaced by a
noninteracting line. 
Finally, it is straightforward to demonstrate using the more 
general criteria introduced by Kadanoff and Baym
\cite{kadbaym} that this theory preserves all conservation laws.

It has been shown  that the results of the full ``mode-coupling''
scheme of  Fig.~\ref{fig:self_e}a are qualitatively captured by the 
lowest order conserving approximation of Fig.~\ref{fig:self_e}(b)
\cite{marcelja}; this approximation, nevertheless, 
goes beyond the original work of Nozi\`eres and Schmitt-Rink. 
Unlike other T-matrix approaches, where higher 
order self-consistency effects tend to diminish leading order
features\cite{Vilk}, it is found \cite{marcelja} that within 
the present framework the inclusion  of ``mode-coupling'' effects
amplify these first order (pseudogap) features. For simplicity, 
we, therefore, focus on the lowest order approximation. The  
discussion of feedback effects is deferred to a  future publication.
\begin{figure}
\narrowtext
\epsfxsize=2.5in 
\hspace{0.3in}\epsfbox{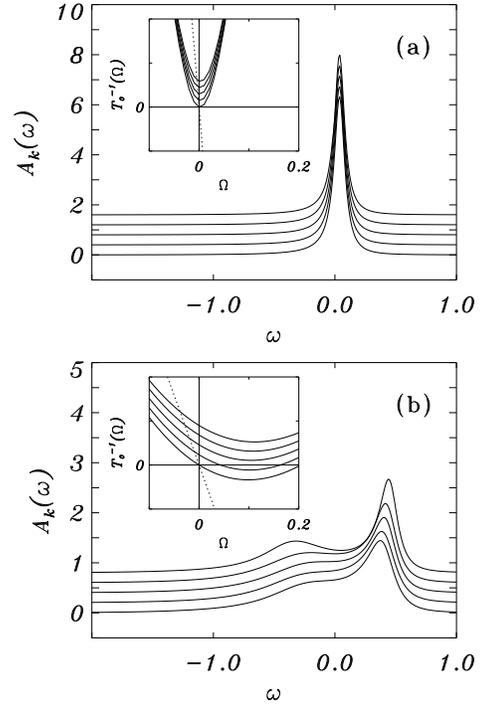}\vspace{0.1in}
\caption{$A_{\bf k}(\omega)$ vs $\omega $ for weak ($g/g_c = 0.6$)  
in (a) and intermediate ($g/g_c = 1.0 $) coupling in (b). $ T/T_c$ 
varies from 1.0 to 1.1. Insets plot ${\rm Re}\, [T^{-1}_{\bf 
q=0}(\Omega )]$ (solid lines, for same $ T/T_c$, as in main figure), 
and ${\rm Im}\, [T^{-1}_{\bf q=0}(\Omega )]$ (dashed lines, at $ T = 
T_c $).}
\label{fig-2}
\end{figure}

In our scenario the  physical process which generates the pseudogap  
is resonant pair scattering (above $T_c$),  arising from the condition
that the real part of the inverse T-matrix,
${\rm Re}\,[T^{-1}_{\bf q} (\Omega = \Omega_{\bf q})] = 0$, when the
imaginary part, ${\rm Im}\,[T^{-1}_{\bf q} (\Omega _{\bf q})]$, is 
sufficiently small.
This resonant behavior is manifested  as a sharp peak in ${\rm
Im}\,[T_{\bf q} (\Omega)]$.  This peak is in turn reflected in the 
electronic self-energy and the spectral function.
The pair resonance condition is illustrated
in the insets of Fig.~\ref{fig-2}, where the behavior of 
${\rm Re}\,[T^{-1}_{\bf q=0}(\Omega)]$, as a function of frequency 
is contrasted for weak ($g/g_c  < 1$) (2a)  and 
intermediate ($g/g_c \approx 1$) (2b)
couplings. Each series of curves corresponds to varying temperature.
The dashed lines indicate  the  form of 
${\rm Im}\, [T^{-1}_{\bf q}(\Omega)]$ at $T_c$ in each of the two cases. The 
critical value $g_c$ establishes the approximate dividing point between  
resonant and non-resonant scattering. 
As can be seen, there is a finite frequency zero 
crossing of  ${\rm Re}\,[T^{-1}_{\bf q}(\Omega)]$
for $T > T_c$, corresponding to resonant scattering,  only in the 
stronger coupling limit. 
The resonance energy increases as a function of temperature $T$ and 
${\bf q}$ until it disappears at a cross-over wave vector ${\bf q}^*$ 
or temperature $T^*$.

The associated spectral functions $A_{\bf k}(\omega)$ for each of the 
two cases considered in the insets are numerically computed from 
the self-energy using Eq.~(\ref{self-energy}) and plotted  for the case 
$k=k_{\rm  F}$ in  the main portion of Fig.~\ref{fig-2}
as a function of $\omega$, for varying $T$. 
(Throughout the unit of energy is $E_F$.) Although the numerical 
integrations involved are computationally intensive, the integrated 
spectral weight is unity to several significant digits 
for each spectral curve presented. 
In the stronger coupling  limit and at sufficiently 
low $T$ (Fig.~\ref{fig-2}(b)), 
the two peaked structure characteristic
of a pseudogap appears and becomes more pronounced with larger
$g/g_c$. In the 
more weakly coupled limit ($g/g_c = 0.6$), the single peak behavior
characteristic of a normal Fermi liquid is recovered, as shown  in
Fig.~\ref{fig-2}(a). In general, the two peaked structure correlates 
with the presence of a resonance in the T--matrix.
For $g$ slightly greater than $g_c$, the two maxima are resolvable
up to $T^*$ of the order of several $T_c$.

\begin{figure}
\narrowtext
\epsfxsize=2.5in 
\hspace{0.3in}\epsfbox{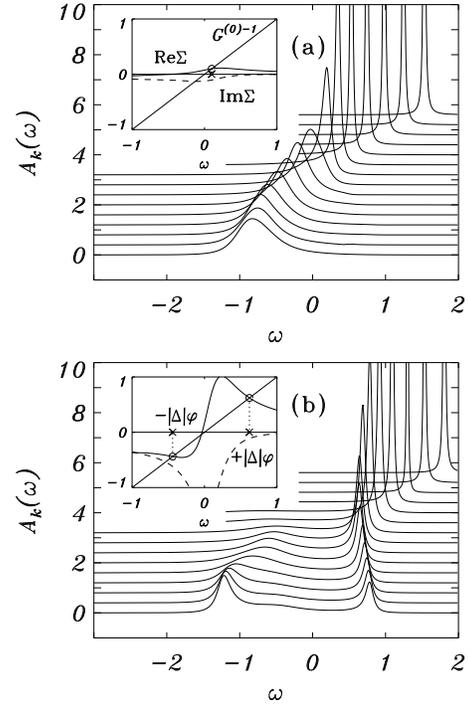}\vspace{0.1in}
\caption{$A_{\bf k} (\omega )$ vs $\omega $ for (a) weak 
($g=0.8g_c$) and (b) intermediate ($g=1.2g_c$) coupling, 
from $k < k_F$ (bottom curve) to $k > k_F$ (top curve). 
The spectra in (b) show signatures of normal
state ``particle-hole mixing''. In the insets are plotted
the corresponding ${\rm Im}\,[\Sigma_{\bf 
k}(\omega)]$ and ${\rm Re}\,[\Sigma_{\bf k}(\omega)]$ 
for $ k = k_F $.} 
\label{fig-3}
\end{figure}

An intuitive understanding of the splitting of the spectral peak 
into a pair of asymmetrically broadened peaks may be gained by 
examining the imaginary part of the self-energy. On the real 
frequency axis ($i\zeta _l \rightarrow \omega + i \delta$), 
${\rm Im}\,[\Sigma_{\bf k}(\omega)]$
is given by
\begin{eqnarray}
{\rm Im}\,\bigl[\Sigma _{\bf k} (\omega)\bigr] = - \sum_{\bf q}
\varphi ^2 _{{\bf k-\bf q}/2}{\rm Im}\bigl[T_{\bf q}(\omega +
\epsilon _{\bf q-k})\bigr] \times \nonumber \\
\bigl[f( \epsilon _{\bf q-k}) + n(\omega + \epsilon _{\bf q-k})\bigr].
\label{imsigma}
\end{eqnarray}
where $f(x),n(x) = (e^{\beta x} \pm 1)^{-1}$ \cite{f-term}.
For intermediate coupling strengths,  a resonance condition leads 
to a peak in ${\rm Im}\,[T_{\bf q}(\Omega)]$
at small frequencies and momenta, which in turn yields a maximum 
in $-{\rm Im}\,[\Sigma_{\bf k}(\omega)]$
at $\omega + \epsilon_{\bf k} \approx 0$ (see the 
inset of Fig. \ref{fig-3}(b)). The frequency weight under this
peak is  written as $\pi|\Delta|^2\varphi^2_{\bf k}$, where  
$|\Delta| $ can be viewed as the pseudogap energy. 
This peak in $-{\rm Im}\,[\Sigma_{\bf k}(\omega)]$ implies -- via the
Kramers--Kr\"onig relation -- a  corresponding resonance structure 
in ${\rm Re}\,[\Sigma_{\bf k}(\omega)]$ at the same frequency 
$\omega \approx - \epsilon _{\bf k}$. In this way  
$A_{\bf k}(\omega)$ acquires two peaks separated by 
$2|\Delta| \varphi_{\bf k}$  with  
\begin{equation}
|\Delta|^2 \approx
 \sum_{\bf q}\int^{+\infty}_{-\infty}\frac{{\rm d}\Omega}{\pi}\,
n(\Omega)\, {\rm Im}\bigr[T_{\bf q}(\Omega)\bigl].
\label{Delta}
\end{equation}
The asymmetric broadening \cite{f-term} of the two
spectral peaks is a generic feature of our results
and is due to the interaction of correlated pairs with
the Fermi sea. This asymmetry, which is contained in Eq.~\ref{imsigma},
reflects that in ${\rm Im}\,[ T_{\bf q} (\Omega )]$, as a function of 
$\Omega $.
In Fig.~\ref{fig-3} we plot  the momentum dependence of the 
spectral 
function slightly above $T_c$ for weak (3a) and intermediate (3b)
coupling, along with typical self--energies shown in the insets.
The former case  shows the single peak structure which 
evolves with ${\bf k}$ in a fashion  characteristic of 
a finite temperature Fermi liquid\cite{f-term}.  
In the stronger coupling limit (Fig.~\ref{fig-3}b) the spectral weight 
shifts from the negative to the positive frequency peak
as the momentum vector ${\bf k}$ passes through the Fermi surface.
Close to the Fermi momentum the peaks disperse roughly as 
$E_{\bf k}=\pm\sqrt{\epsilon_{\bf k}^2+|\Delta|^2\varphi^2_{\bf k}}$.
This dispersion provides a predictive 
signature for future ARPES measurements, within  the precursor 
superconductivity scenario. Indeed,  this behavior is reminiscent of 
the particle-hole mixing found in photoemission measurements on 
the superconducting state\cite{Ding,Lee}. 

\begin{figure}
\narrowtext
\epsfxsize=2.5in 
\hspace{0.5in}\epsfbox{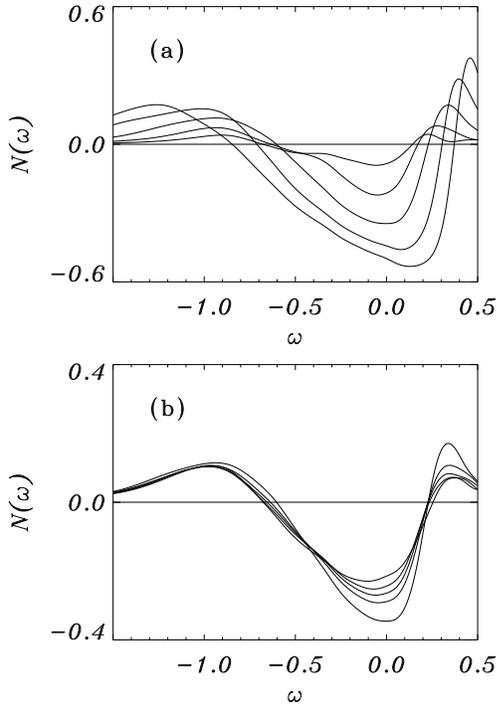}\vspace{0.1in}
\caption{$N (\omega )$ vs $\omega$ for $g/g_c = 0.8 $ to
$ 1.1$. and $T= T_c$ (a), and for $T/T_c = 1.0 $ to $1.2$, at $ 
g = g_c$, (b). The normal state density of states is subtracted off in
both cases (see text).}
\label{fig-4}
\end{figure}

Finally,  the density of states, $N(\omega)$, is plotted
in Fig.~\ref{fig-4} as a function of energy. 
This quantity may be directly related 
to tunneling as well as to thermodynamic measurements in the 
pseudogap regime. The asymmetry in the curves reflects, in part, the 
asymmetry of the spectral functions seen in Fig.~\ref{fig-2} and 
\ref{fig-3}. 
For clarity the results are represented by subtracting the
``normal'' state curve, obtained, for 
definiteness, in the very weak coupling limit. Fig.~\ref{fig-4}(a)
indicates the coupling constant dependence of $N(\omega)$
and Fig.~\ref{fig-4}(b) the corresponding temperature dependence 
for fixed $g$. A depression in $N(\omega )$ -- which increases 
with $g$ -- develops at smaller couplings, and persists to 
higher temperatures, than do pseudogap effects
in the spectral function (see Fig.~\ref{fig-3}). 

In summary, we have demonstrated how resonant pair scattering 
above
$T_c$ gives rise to a splitting of the spectral function, $A_{\bf 
k}(\omega)$, as well as a density of states depression. 
Experimental observation of the former is the 
more significant manifestation of pseudogap behavior, providing 
strong constraints on theoretical models. Our precursor 
superconductivity scenario has predictive signatures: an 
asymmetry in the widths of the two spectral peaks
and a ${\bf k}$-dependent dispersion of the $T > T_c $ spectral 
function, qualitatively similar to that of the BCS state.
A $d_{x^2 - y^2 }$ symmetry of the normal state gap will arise 
naturally in  the present scenario, for a $d$--wave superconducting 
instability. This would be accompanied by  a spectral peak broadening 
proportional to $(\cos k_x - \cos k_y)^2$ . The present picture
should  be differentiated from preformed pair models: 
the correlated pairs of our picture have significant spatial extent and
fail to obey Bose statistics. Furthermore, in contrast to
the stripe picture of Emery and Kivelson, the amplitude and 
phase of this paired state is never established beyond the 
dimensions and lifetime of a single pair. Quasi two dimensionality 
will enhance our pseudogap effects, which should, then,
become more pronounced as the insulator is approached. 
Magnetic correlations may, also, ultimately  play 
a role in the extreme underdoped regime. Nevertheless, short 
coherence lengths and quasi 2d features suggest that
precursor superconductivity is present to some degree and must 
necessarily be calibrated in order to obtain a full understanding 
of the cuprate pseudogap. 

We would like to thank J. C. Campuzano, H. Ding, L. P. Kadanoff,
A. Klein, I. Kosztin, P. C. Martin, M. Norman, B. R. Patton, 
and especially Y. M. Vilk for useful discussion. 
This research was supported in part by the Natural Sciences
and Engineering Research Council of Canada fellowship (J. M.) and
the Science and Technology Center for Superconductivity funded by 
the National Science Foundation under award No. DMR 91--20000.

\end{document}